# Tracing the origin of a scientific legend by Reference Publication Year Spectroscopy (RPYS): the legend of the Darwin finches

Werner Marx

Max Planck Institute for Solid State Research, Heisenbergstr. 1, D-70569 Stuttgart, Germany; w.marx@fkf.mpg.de

Lutz Bornmann

Division for Science and Innovation Studies, Administrative Headquarters of the Max Planck Society, Hofgartenstr. 8, 80539 Munich, Germany; bornmann@gv.mpg.de

**Abstract**

In a previews paper we introduced the quantitative method named Reference Publication Year Spectroscopy (RPYS). With this method one can determine the historical roots of research fields and quantify their impact on current research. RPYS is based on the analysis of the frequency with which references are cited in the publications of a specific research field in terms of the publication years of these cited references. In this study, we illustrate that RPYS can also be used to reveal the origin of scientific legends. We selected "Darwin finches" as an example for illustration. Charles Darwin, the originator of evolutionary theory, was given credit for finches he did not see and for observations and insights about the finches he never made. We have shown that a book published in 1947 is the most-highly cited early reference cited within the relevant literature. This book had already been revealed as the origin of the term "Darwin finches" by Sulloway through careful historical analysis.

**Introduction**

Research activity usually evolves on the basis of previous investigations and discussions among the experts in a scientific community. Although there are many differences between the theories of scientific development, the relationship of current research to past literature plays a significant role: knowledge cannot be acquired without this relationship. The relationship to earlier publications is expressed in the form of references. One can expect that the content of the cited publication and that of the citing publication are related and that the former is of significance to the knowledge claim in the latter. The premise of a normative theory of citations is that the more frequently scientific publications are cited, the more important they are for the advancement of knowledge. From this perspective, citation data provides interesting insight into the historical science context, in terms of the significance of the previous historical publications on which the later publications in a field are based. Garfield discussed the application of citation data to sociological and historical research as early as 1963 [1].

**Reference Publication Year Spectroscopy (RPYS)**

In a previous publication, rather than starting with citations of the publications in a specific research field, for certain issues we proposed reversing the perspective and analyzing the papers referenced in the publications in the same research field in order to determine the impact of publications, authors, institutions or journals within that field [2]. A cited reference analysis with specific emphasis on the publication years of the references can be used to quantify the significance of historical publications and to reveal the historical roots of a given research field. The analysis of the publication years of references is not a new bibliometric approach but has already been discussed by De Solla Price (De Solla Price, 1974; p. 91). And it was for instance applied by Van Raan to measure the growth of science and to detect important breakthroughs in science without pre-defining any fields (van Raan, 2000).

Quantitative analysis of the reference publication years (RPYs, not to confuse with the method RPYS) in a specific research field shows that frequently occurring RPYs become more differentiated towards the past and mostly show up as distinct peaks. If one analyzes the publications underlying these peaks, one will find that during the 19th and the first half of the 20th century they are predominantly formed by single relatively highly cited publications. These few, particularly frequently cited publications contain as a rule the historic papers (or even the historical roots) most relevant for the evolution of a specific research field which should be taken into consideration when discussing its history. Their specific role can only be revealed by careful analysis through experts in the relevant field. In an analogy to the spectra in signal analysis, which are characterized by pronounced peaks in the quantification of certain properties (such as the absorption or reflection of light as a function of its color), we call this special application Reference Publication Year Spectroscopy (RPYS) [3-4].

RPYS can be used to analyze the historical roots of a research field, but can also be used to reveal the origin of a scientific legend. This study deals with the origin of the term "Darwin finches", which has been introduced into scientific literature and beyond as a legend and is based on a misconception. The legend implies that the finches of the Galapagos Islands prompted Darwin to develop his theory on the origin of species based on the mechanism of natural selection. On principle, this could have happened but definitely did not: In the first edition (1839) of his Journal of Researches Darwin [5] said very little about the finches and in his book on the origin of species [6] they were not even mentioned. The legend was obviously imposed on the history of the theory of evolution later, giving rise to the question of which publication this term comes from. Since the answer has already been given by an expert of the theory of evolution [7-8], we can verify the RPYS method and will show that the roots of a legend (here the "Darwin finches") can also be revealed by using this bibliometric method. Scientific legends have long been a topic in popular as well as in scientific literature (see for example references [9], [10], and [11]). Wetterer has already applied bibliometric data to the tracing of a scientific legend [12].

**Methods**

The result of the application of RPYS to literature dealing with "Darwin finches" presented here is based on the Science Citation Index (SCI) which is accessed via the SCIsearch database offered by the database provider STN International (http://www.stn-international.com/). This database combined with the STN search system enables sophisticated citation analysis. Among many other options, the SCIsearch database searched via STN makes it possible to ask which historical publications in the various fields of the natural sciences have been cited most frequently by the papers published since 1974, the period covered by the SCIsearch database. The Web of Science (WoS) provided by Thomson Reuters, the most common search platform of the Thomson Reuters citation indexes, stretches back to 1900. However, the WoS search functions have not been optimized for the kind of bibliometric analysis presented here.

STN's retrieval system allows the publications from a specific research field to be selected and all the references they cite to be extracted. Instead of the complete references it is also possible to select and analyze only the authors of the publications in the cited references, only the journals or only the RPYs. In this study, we are concerned mainly with the analysis of the RPYs and especially the early publications cited particularly frequently as the origins of specific terms (here "Darwin finches"). The first step in RPYS is to select the publications for a certain research field and extract all the references from them. The second step is to establish the distribution of the frequencies of the cited references over the RPYs and from this determine the early RPYs cited most frequently (a minimum citation count of 10 has proved to be reasonable). The third is to analyze these RPYs for frequently cited historical publications.

The publications dealing with "Darwin finches" were selected in the SCI database by searching with the term "Darwin('s) finch(es)". All cited references with reference publication years prior or equal to 1960 (n=

1961) have been selected from the references of the complete set of 689 papers on "Darwin finches" published within scientific literature since 1974.

## Results

Figure 1 shows the distribution of the cited references limiting the reference publication years to the time period 1800 to 1960. The reference publication years stretch back to before 1800. However, a large number of references with early reference publication years usually prove to be erroneous. Whereas Figure 1a shows the distribution of the number of cited references across the publication years, Figure 1b shows the absolute deviation of the number of cited references in one year from the median for the number of cited references in the two previous, the current and the two following years. The deviation from the median makes the distinct peaks more clearly visible in a historical context.

Figures 1a and 1b show some distinct peaks. In a historical context, the vast majority of the cited references collected under a distinct peak results from the citations of one single publication.

The first distinct peak in Figure 1 can be assigned to the reference publication year 1859: 53 out of altogether 54 citations refer to Charles Darwin's book "On the Origin of Species by Means of Natural Selection, or the Preservation of Favoured Races in the Struggle for Life" first published in 1959 [6]. The next peak can be assigned to Darwin's second book on evolutionary theory, following his 1859 work: 21 out of 24 citations refer to Darwin's book "The Descent of Man, and Selection in Relation to Sex" first published in 1871 [13]. The next distinct peak can be assigned to a book by Theodosius Dobzhansky published in 1937 [14]: 22 out of 41 citations refer to his book entitled "Genetics and the Origin of Species", a major work on the modern evolutionary synthesis (the synthesis of evolutionary biology with genetics). Due to their overall importance, it is not surprising that these basic works on evolutionary theory are frequently cited by papers dealing with "Darwin finches". However, a careful reading of these books shows that they are not the origin of the legend about Darwin finches.

The most clearly pronounced peak in Figure 1 can be assigned to the reference publication year 1947: 144 out of 161 citations refer to a book by D.L. Lack entitled "Darwin's finches" published in 1947 [15]. The strong suspicion that this book is the origin of the term "Darwin finches" is confirmed by the historical analysis of F.J. Sulloway published in 1982 in his paper "Darwin and his finches: The evolution of a legend" [7]. A short form of this paper was also published in *Nature* [8]. Although David Lack claimed that the term "Darwin finches" was first applied by Percy Lowe in 1936 (there is also a corresponding peak visible in Figure 1), it was only then popularized by Lack in his 1947 book. The following two quotations from the paper by Sulloway [7] illustrate the emergence of the legend originated by the book by David Lack published in 1947:

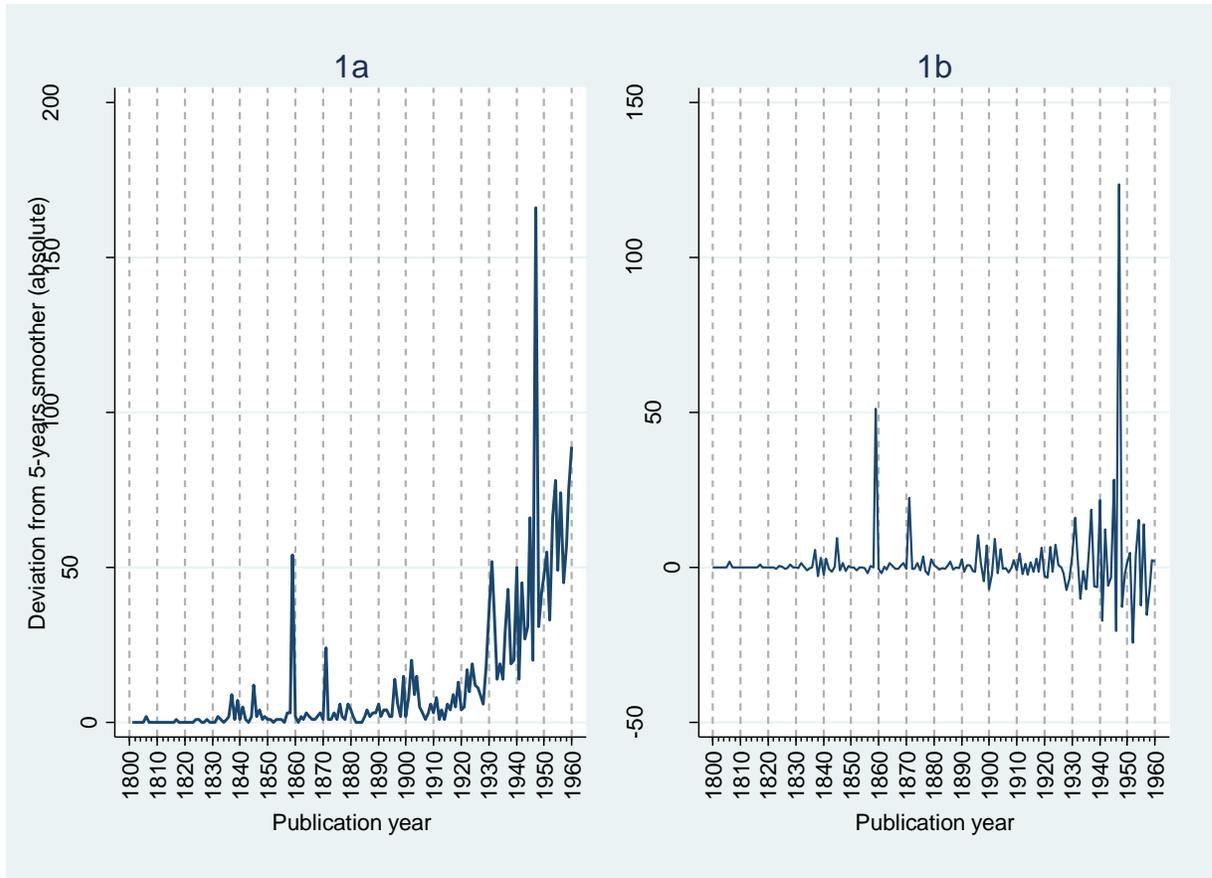

**Figure 1:** Distribution of the cited references across the reference publication years 1800 to 1960: Whereas Figure 1a shows the distribution of the number of cited references across the publication years, Figure 1b shows the absolute deviation of the number of cited references in one year from the median for the number of cited references in the two previous, the current and the two following years.

(1) "As it turns out, Darwin made absolutely no effort while in the Galapagos to separate his finches by island; and what locality information he later published, he reconstructed after his return to England, using other shipmates' carefully labeled collections ... Even after his return to England, when John Gould had clarified the affinities of this unusual avian group, Darwin was slow to understand how the Galapagos finches had evolved ... Lastly, far from being crucial to his evolutionary argument, as the legend would have us believe, the finches were not even mentioned by Darwin in the Origin of Species. In spite of the legend's manifest contradictions with historical fact, it successfully holds sway today in the major textbooks of biology and ornithology, and is frequently encountered as well in the historical literature on Darwin. It has become, in fact, one of the most widely circulated legends in the history of the life sciences, ranking with the famous stories of Newton and the apple and of Galileo's experiments at the Leaning Tower of Pisa, as a classic textbook account of the origins of modern science" (pp. 39-40).

(2) "Although Lack was not the first person to use this term, it was he who succeeded in popularizing it. In one sense the term is felicitous, because not all the Geospizinae [scientific name of Darwin finches] are confined to the Galapagos Islands, and thus the name 'Galapagos finches' is inappropriate for the whole group. This was, in fact, the chief reason for Lack's use of the expression 'Darwin's finches.' But as the term became more popularly known through Lack's book, people tended to assume that these birds had been so named because, as one biologist put it, 'they helped to persuade Darwin of the truth of evolution' and were crucial as well to his later theories. Through this act of eponomy, Darwin was increasingly given credit after 1947 for finches he never saw and for observations and insights about them he never made" (pp. 45-46).

**Discussion**

The origin of the term "Darwin finches" was taken as a case study to point out the potential of the RPYS method [3-4] to investigate scientific legends. We have selected the literature on "Darwin finches" and have applied RPYS to identify possible publications as the origin of this legend which has spread out into the scientific literature and beyond. We have shown that the book by David Lack published in 1947 [15] is by far the most-highly-cited pre-1960 reference cited since 1974 within the relevant literature. This book has been already identified as the origin of the misleading term "Darwin finches" by the extensive historical analysis of Sulloway [7-8]. This study has shown that RPYS (a simple analysis of cited references in a publication set) has the potential to make the origin of a legend visible in a similar way to a sophisticated and time-consuming historical analysis. However, we should consider that the method is only as good as the publication-citation record. As illustrated by our example, one ends up with the paper that popularized the concept (here: the legend) under study, not necessary the paper that introduced the concept. If everyone has forgotten the original source of a concept, then only careful historical analysis can reveal the true origin.